\documentclass[12pt]{article}
\usepackage[dvips]{graphicx}
\begin{document}
\centerline{\large\bf $CP$ violation in long baseline neutrino}
\vskip 0.6truecm
\centerline{\large\bf oscillation experiments}
\baselineskip=8truemm
\vskip 1.0truecm
\centerline{
Chise Sakaguchi$,^{a),}$\footnote{e-mail: chise-s@par.odn.ne.jp}
\ Toshihiko Hattori$,^{a),}$\footnote{e-mail: hattori@ias.tokushima-u.ac.jp} \ and
\ Seiichi Wakaizumi$,^{b),}$\footnote{e-mail: wakaizum@medsci.tokushima-u.ac.jp}
}
\vskip 0.6truecm
\centerline{
\it $ ^{a)}$Institute of Theoretical Physics, University of Tokushima, Tokushima 770-8502, Japan}
\centerline{
\it $ ^{b)}$Center for University Extension, University of Tokushima, Tokushima 770-8502, Japan}
\vskip 1.0truecm
\centerline{\bf Abstract}
\vskip 0.7truecm

We define a difference $D_{\rm CP}$ of the neutrino oscillation probability differences 
with matter effect for the CP-conjugate channels, divided by neutrino beam energy, 
taken between the two baselines $L = L_1$ and $L =L_2$ with $L_1/E_1 = L_2/E_2$,
where $E_1$ and $E_2$ are the neutrino energy
for the experiment with $L_1$ and $L_2$, respectively.
The quantity $D_{\rm CP}$ doesn't contain the matter effect 
to the first order in $aL/2E$, $a$ representing the matter effect.
We show  the behavior of $D_{\rm CP}$
with $L_1 = 300$ km fixed and $L_2$ variable in the three-neutrino model.

\newpage

Where does $CP$ violation originate?
In order to study the origin of $CP$ violation, we expect that
the observation of $CP$ violation in neutrino oscillation experiments will be fruitful.

The neutrino oscillation is a strong means
to examine the masses and mixing angles of the neutrinos~\cite{Maki}.
The experiments have shown the solar neutrino deficit~\cite{Solar} 
and the atmospheric neutrino anomaly~\cite{Hirata},
which strongly indicate the neutrino oscillation~\cite{Fogli}.
The large mixing angle solution (LMA) by means of MSW effect~\cite{Wolfenstein} 
to the solar neutrino problem gives a mass-squared difference of
$10^{-5}-10^{-4} {\rm eV}^{2}$ ~\cite{Bahcall},
and the atmospheric neutrino anomaly brings the mass-squared difference of 
$(1.5-5)\times 10^{-3} {\rm eV}^{2}$~\cite{Fukuda}.
Especially, long baseline neutrino oscillation ecperiments are planned~\cite{Itow} 
to measure precisely the mass-squared differences and the mixing angles and, 
moreover, the CP violation effects in the neutrino oscillation~\cite{Cabibbo}.
For the long baseline experiments, however, the matter effect gives a fake CP violation effect 
comparable to the pure CP violation effect~\cite{Barger,Arafune}.
Therefore, it is necessary to know
how to distinguish the pure CP violation effect from the matter effect.

In this paper we will study the behavior of pure CP violation effects
with the quantity $D_{\rm CP}$ (difference of the CP violation effects) newly introduced.

We assume three generations of neutrinos
which have mass eigenstates $\nu^{'}_{i}$ with mass $m_{i}(i=1,2,3)$.
The flavor eigenstates $ \nu_{\alpha} (\alpha=e,\mu,\tau)$ and the mass eigenstates in the vacuum are related as
    \begin{equation}
    \nu_{\alpha}=U^{(0)}_{\alpha i}\nu^{'}_{i}
    \end{equation}
by mixing matrix $U^{(0)}$. We take
    \begin{equation}
    U^{(0)}=\pmatrix{
    c_{\phi}c_{\omega} 
    & c_{\phi}s_{\omega}
    & s_{\phi} \cr
    -c_{\psi} c_{\omega}- s_{\psi} s_{\phi}c_{\omega} e^{i\delta}
    & c_{\psi} c_{\omega}- s_{\psi} s_{\phi}s_{\omega} e^{i\delta}
    & s_{\psi} c_{\phi} e^{i\delta} \cr
    s_{\psi} s_{\omega}- c_{\psi} s_{\phi}c_{\omega} e^{i\delta}
    & -s_{\psi} c_{\omega}- c_{\psi} s_{\phi}s_{\omega} e^{i\delta}
    & c_{\psi}c_{\phi} e^{i\delta} \cr
    }
    \end{equation}
as mixing matrix $U^{(0)}$, where $c_{\psi}=\cos\psi $, $s_{\psi}=\sin\psi $, etc.

According to Arafune, Koike and Sato's formalism~\cite{Arafune},
the evolution equation for the flavor eigenstate vector in the vacuum is expressed as
    \begin{equation}
    i \frac{d \nu}{dx}
    =\frac{1}{2E} U^{(0)} diag(0,\delta m^{2}_{21},\delta m^{2}_{31}) U^{(0)\dag}\nu
    \end{equation}
where $E$ is the energy and $ \delta m_{ij}^{2}=m_{i}^{2}-m_{j}^{2}$.
Similarly the evolution equation in matter is given as
    \begin{equation}
    i\frac{d \nu}{dx}=H \nu,
    \end{equation}
where 
    \begin{equation}
    H \equiv \frac{1}{2E} U diag(\mu_{1}^{2} , \mu_{2}^{2} , \mu_{3}^{2} ) U^{\dag}.
    \end{equation} 
A unitary mixing matrix $U$ and the effective mass squared $\mu_{i}^{2} (i=1,2,3)$ are determined by
    \begin{equation}
    U \pmatrix{
    \mu_{1}^{2} & 0 &0 \cr 
    0 & \mu_{2}^{2} & 0 \cr 
    0 & 0 & \mu_{3}^{2} \cr}
    U^{\dag}
    = U^{(0)}\pmatrix{
    0 & 0 &0 \cr 
    0 & \delta m_{21}^{2} & 0 \cr 
    0 & 0 & \delta m_{31}^{2} \cr} 
    U^{(0)\dag}
    +\pmatrix{a&0&0 \cr  0&0&0 \cr  0&0&0 \cr},
    \end{equation}
with
    \begin{equation}
    a \equiv 2\sqrt{2} G_{F} n_{e} E
    =7.56\times10^{-5} {\rm eV}^{2} \frac{\rho}{{\rm g cm}^{-3}} \frac{E}{\rm GeV}, 
    \end{equation}
where $n_{e}$ is the electron density and $ \rho $ is the matter density.

The solution of Eq. (4) is
    \begin{equation}
    \nu (x)=S(x) \nu (0),
    \end{equation}
where 
    \begin{equation}
    S \equiv T e^{-i\int ^{x}_{0}ds H(s)}
    \end{equation}
and $T$ is the symbol for time ordering. 
$S$ gives the oscillation probability 
for $ \nu_{\alpha}\rightarrow\nu_{\beta} (\alpha,\beta=e,\mu,\tau)$ at distance $L$ as
    \begin{equation}
    P(\nu_{\alpha}\rightarrow\nu_{\beta};L)=\bigl| S_{\beta\alpha}(L) \bigr|^{2}.
    \end{equation} 
The oscillation probability for the antineutrino
$P(\overline{\nu}_{\alpha} \rightarrow \overline{\nu}_{\beta};L)$ is obtained by replacing 
$a\rightarrow -a $ and $U \rightarrow U^{*}$ in Eq.(10).

Taking Arafune et al.'s formalism~\cite{Arafune} in order to calculate Eq.(10) up to the first order in $aL/2E$,
we then obtain the oscillation probability 
$P(\nu_{e }\rightarrow\nu_{\tau})$
in the lowest order approximation as
    \begin{eqnarray}
    P(\nu_{e} \rightarrow \nu_{\tau})
      &=&4 \sin^{2} \frac{\delta m^{2}_{31} L}{4E} c^{2}_{\phi} s^{2}_{\phi} c^{2}_{\psi}
         \left[ 1-2 \frac{a}{\delta m^{2}_{31}} (2 s^{2}_{\phi}-1)  \right] \nonumber \\
      & & {}+2\frac{\delta m^{2}_{31} L}{2E}\sin\frac{\delta m^{2}_{31} L}{2E}
         c^{2}_{\phi}s_{\phi}c_{\psi}  \nonumber \\
      & & {}\times 
         \left[ 
         -\frac{a}{\delta m^{2}_{31}} s_{\phi}c_{\psi}(1-2s^{2}_{\phi})
         +\frac{\delta m^{2}_{21}}{\delta m^{2}_{31}} s_{\omega}
         (-s_{\phi}c_{\psi}s_{\omega}-s_{\psi}c_{\omega}c_{\delta})
         \right]  \nonumber \\
      & & {}-4\frac{\delta m^{2}_{21}}{2E} \sin^{2}\frac{\delta m^{2}_{31}}{4E}
         c^{2}_{\phi} s_{\phi} c_{\psi} s_{\psi} c_{\omega} s_{\omega} s_{\delta},
    \end{eqnarray}
and
$P(\nu_{\mu}\rightarrow\nu_{e})$,
$P(\nu_{\mu}\rightarrow\nu_{\mu})$
and $P(\nu_{\mu}\rightarrow\nu_{\tau})$
are given in Arafune et al.'s paper~\cite{Arafune}.
Recalling that $P(\overline{\nu}_{\alpha} \rightarrow \overline{\nu}_{\beta})$ is obtained from 
$P(\nu_{\alpha} \rightarrow \nu_{\beta})$ by the replacement
$ a\rightarrow -a $ and $ \delta \rightarrow -\delta $ ,we define
    \begin{equation}
    \Delta P(\nu_{\alpha}\rightarrow\nu_{\beta})
    \equiv P(\nu_{\alpha} \rightarrow \nu_{\beta})
          -P(\overline{\nu}_{\alpha} \rightarrow \overline{\nu}_{\beta}).
          \end{equation}
Then we have  
    \begin{eqnarray}
    \Delta P(\nu_{\mu}\rightarrow\nu_{e})
        &=&16\frac{a}{\delta m^{2}_{31}} 
          \left[ 
          \sin^{2}\frac{\delta m^{2}_{31} L}{4E}
          -\frac{1}{4} \frac{\delta m^{2}_{31}L}{2E}\sin\frac{\delta m^{2}_{31}L}{2E}
          \right] \nonumber \\
        & & {} \times c^{2}_{\phi} s^{2}_{\phi} s^{2}_{\psi}(1-2s^{2}_{\phi}) \nonumber \\
        & & {}-8\frac{\delta m^{2}_{21}L}{2E}\sin^{2}\frac{\delta m^{2}_{31} L}{4E}
          c^{2}_{\phi} s_{\phi} c_{\psi} s_{\psi} c_{\omega} s_{\omega} s_{\delta},
    \end{eqnarray}
    \begin{eqnarray}
    \Delta P(\nu_{\mu}\rightarrow\nu_{\mu })
        &=&16\frac{a}{\delta m^{2}_{31}} 
           \left[ 
           \sin^{2}\frac{\delta m^{2}_{31} L}{4E}
           -\frac{1}{4} \frac{\delta m^{2}_{31}L}{2E}\sin\frac{\delta m^{2}_{31}L}{2E}
           \right] \nonumber \\
        & & {} \times c^{2}_{\phi} s^{2}_{\phi} s^{2}_{\psi}(1-2c^{2}_{\phi} s^{2}_{\psi}),
    \end{eqnarray}
    \begin{eqnarray}
    \Delta P(\nu_{\mu}\rightarrow\nu_{\tau})
      &=&16\frac{a}{\delta m^{2}_{31}} 
        \left[ 
        \sin^{2}\frac{\delta m^{2}_{31} L}{4E}
        -\frac{1}{4} \frac{\delta m^{2}_{31}L}{2E}\sin\frac{\delta m^{2}_{31}L}{2E}
        \right] \nonumber \\
      & & {} \times c^{2}_{\phi} s^{2}_{\phi} s^{2}_{\psi}(-2c^{2}_{\phi} c^{2}_{\psi}) \nonumber \\
      & & {}+8\frac{\delta m^{2}_{21}L}{2E}\sin^{2}\frac{\delta m^{2}_{31} L}{4E}
        c^{2}_{\phi} s_{\phi} c_{\psi} s_{\psi} c_{\omega} s_{\omega} s_{\delta},
    \end{eqnarray}
    \begin{eqnarray}
    \Delta P(\nu_{e}\rightarrow\nu_{\tau})
            &=&16\frac{a}{\delta m^{2}_{31}} 
               \left[ 
               \sin^{2}\frac{\delta m^{2}_{31} L}{4E}
               -\frac{1}{4} \frac{\delta m^{2}_{31}L}{2E}\sin\frac{\delta m^{2}_{31}L}{2E}
               \right] \nonumber \\
            & & {} \times c^{2}_{\phi} s^{2}_{\phi} c^{2}_{\psi}(1-2s^{2}_{\phi}) \nonumber \\
            & & {}-8\frac{\delta m^{2}_{21}L}{2E}\sin^{2}\frac{\delta m^{2}_{31} L}{4E}
               c^{2}_{\phi} s_{\phi} c_{\psi} s_{\psi} c_{\omega} s_{\omega} s_{\delta},
    \end{eqnarray}
As $\Delta P(\nu_{\mu}\rightarrow\nu_{\mu})$ is independent of $ \delta $,
we see that it doesn't give the pure-$CP$ violation effect 
and consists only of the matter effect term.

Now we separate out the pure $CP$-violation effect from the net $CP$-violation 
by means of the results of experiments with two different baseline $L$ 's.
Suppose that two experiments with $L=L_{1}$ and $L=L_{2}$ are available. 
We observe two probabilities $P(\nu_{\alpha}\rightarrow\nu_{\beta} ;L_{1})$
at neutrino energy $E_{1}$  and $P(\nu_{\alpha} \rightarrow\nu_{\beta} ;L_{2})$ 
at energy $E_{2}$ with $L_{1}/E_{1}= L_{2}/E_{2}(\alpha\ne\beta)$.
Because the matter effect factor $a$ is proportional to energy $E$,
we obtain the matter effect as a function of $L/E$
with dividing $\Delta P(\nu_{\alpha} \rightarrow\nu_{\beta} )$
by energy $E$ in each experiment.
And we define the difference $D_{\rm CP}$ as
    \begin{equation}
    D_{\rm CP} \equiv
    \left[
    \frac{1}{E_{1}}\Delta P(L_{1})-\frac{1}{E_{2}}\Delta P(L_{2})
    \right]_{\frac{L_{1}}{E_{1}}=\frac{L_{2}}{E_{2}}}.
    \end{equation}

\noindent
The quantity $D_{\rm CP}$ contains no matter effect to the first order in $aL/2E$.
We note that this quantity is different from the one defined by Arafune et al\cite{Arafune}.
In Figs.1-3 we show $D_{\rm CP}$ by taking $\Delta P(L)$'s with two different baselines.
In Figs.1 and 2 we show $D_{\rm CP}$ 
for $L_1 =$ 300 {\rm km}, $L_2 =$ 50 {\rm km}
and $L_1 =$ 300 {\rm km}, $L_2 =$100 {\rm km}, respectively.
We have taken
$\Delta m^2_{32} \equiv \Delta m^2_{\rm atm} = 2.5\times 10^{-3} {\rm eV}^{2},
\Delta m^2_{21} \equiv \Delta m^2_{\rm solar} = 4.9\times 10^{-5} {\rm eV}^{2}$,
and the mixing angles and phases as
$s_\omega = 0.53, s_\psi = 0.74, s_\phi = 0.16$ and $\delta = \pi/2$.
Since $D_{\rm CP}$ does not involve the matter effect,
we have used the exact expressions of
$\Delta P(L)$ for the pure {\it CP}-violation effects in the computation of $D_{\rm CP}$. 
As can be seen in Figs.1 and 2, there are two large peaks in $D_{\rm CP}$
around $E = 0.12$ GeV and 0.2 GeV at $ L=300 {\rm km}$.
The peaks become smaller, as the second baseline increases.
In Fig.3 we compare the magnitude of $D_{\rm CP}$ for various values of $L_2$
with $L_1$ fixed as 300 km.

Finally, as the quantity $D_{\rm CP}$ does not involve the matter effect
to the first order in $aL/2E$, it is not affected by the matter effect up to the order of about 5\%
for $\delta (D_{\rm CP})/D_{\rm CP}$ for $\rho = 3$ ${\rm g/cm}^{3}$ and $L = 300$ {\rm km}.
If $\Delta P(L)$ is measured to the accuracy of 10\% $(\delta(\Delta P)/\Delta P \sim 0.1)$
and the neutrino beam energy is focussed to the precision of 10\% $(\delta E/E \sim 0.1)$,
then the quantity $D_{\rm CP}$ will be observed 
to the accuracy of 20\% $(\delta (D_{\rm CP})/D_{\rm CP}\sim 0.2)$.
We hope that $D_{\rm CP}$ will be measured in the future.

\newpage

\begin{figure}
\begin{center}
\includegraphics[width=14cm]{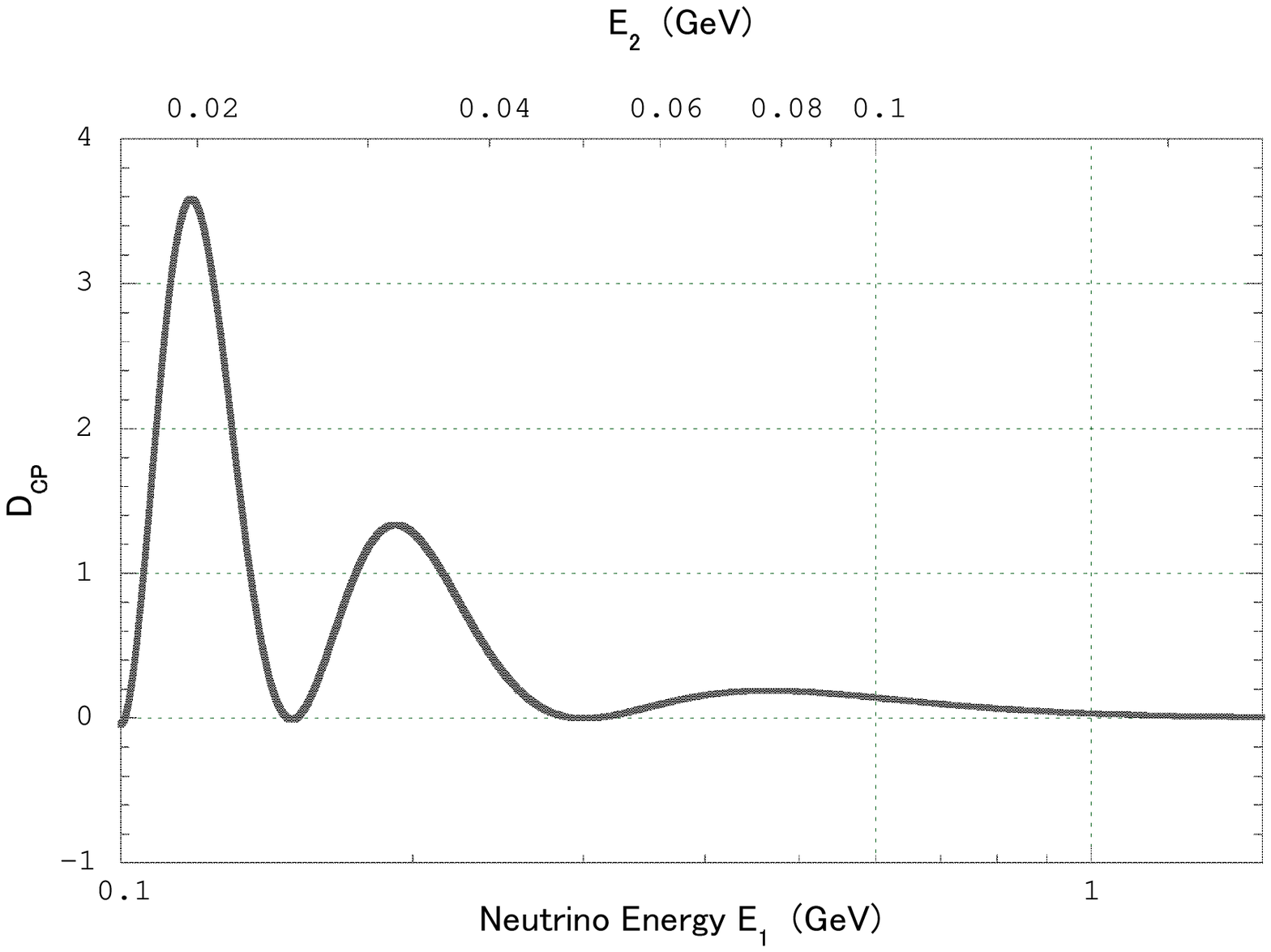}
\end{center}
\caption {
The difference $D_{\rm CP}$ for $L_1=300 {\rm km}$ and $L_2=50{\rm km}$.
$E_1$ and $E_2$ are the neutrino energy for $L_1$ and $L_2$, respectively.
}
\end{figure}
\vskip 0.5truecm

\newpage

\begin{figure}
\begin{center}
\includegraphics[width=14cm]{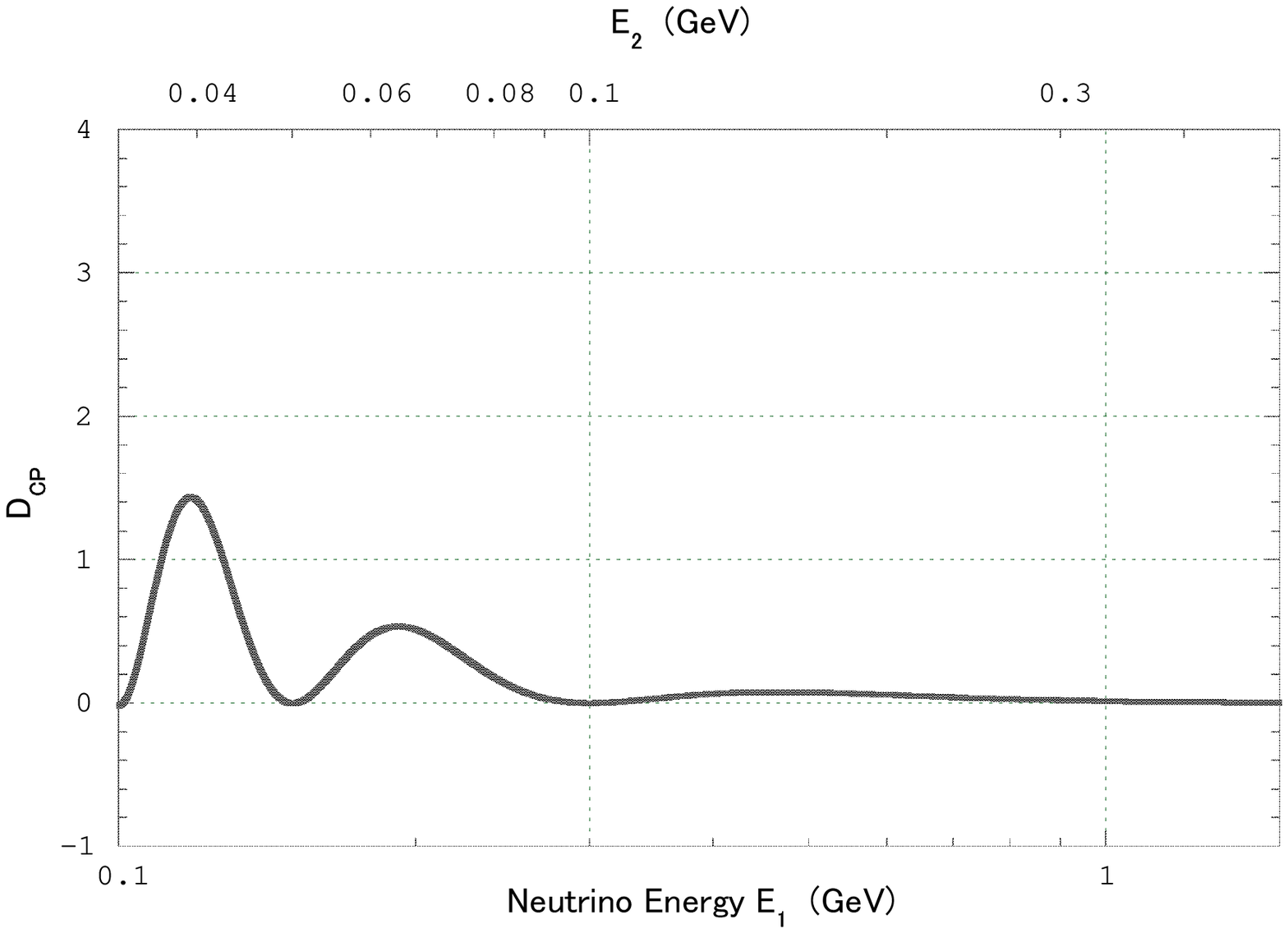}
\end{center}
\caption {
The difference $D_{\rm CP}$ for $L_1=300 {\rm km}$ and $L_2=100{\rm km}$.
$E_1$ and $E_2$ are the neutrino energy for $L_1$ and $L_2$, respectively.
}
\end{figure}
\vskip 0.5truecm

\newpage

\begin{figure}
\begin{center}
\includegraphics[width=14cm]{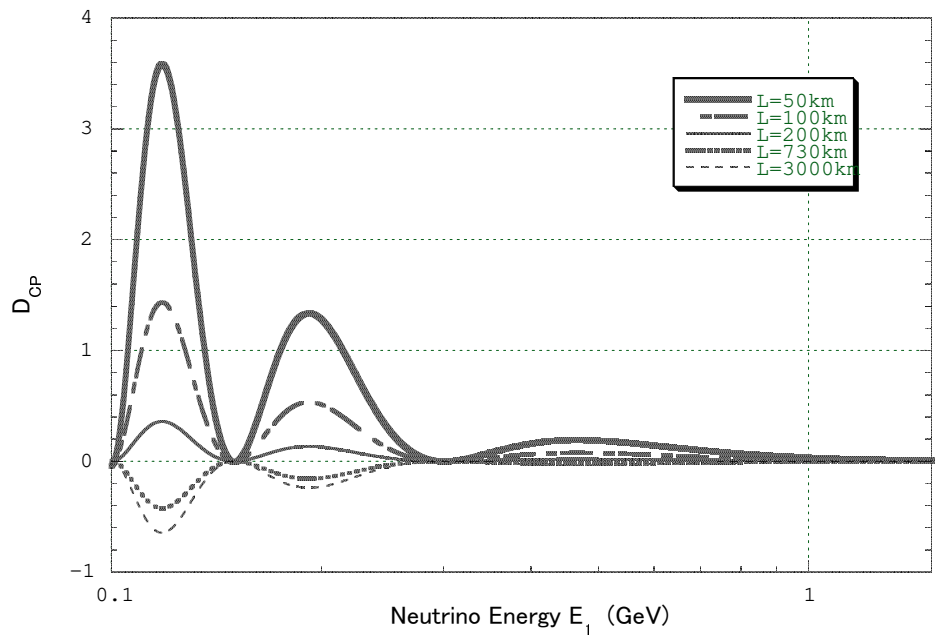}
\end{center}
\caption {
The difference $D_{\rm CP}$ for several values of $L_2$ 
with $L_1=300 {\rm km}$ fixed.
}
\end{figure}
\vskip 0.5truecm

\end{document}